\documentclass[reprint,amsmath,amssymb,aps,pra,longbibliography]{revtex4-1}

\usepackage{graphicx}
\usepackage{dcolumn}
\usepackage{bm}
\usepackage{bbm}
\usepackage{todonotes}

\DeclareSymbolFont{Eulerscripteusm10}{U}{eus}{m}{n}
\SetSymbolFont{Eulerscripteusm10}{bold}{U}{eus}{b}{n}
\DeclareMathSymbol{\euW}{\mathord}{Eulerscripteusm10}{"57}
\DeclareMathSymbol{\euD}{\mathord}{Eulerscripteusm10}{"44}

\DeclareMathAlphabet{\mathbxit}{\encodingdefault}{\rmdefault}{bx}{it}   

\newcommand{\be}{\begin{eqnarray}}
\newcommand{\ee}{\end{eqnarray}}
\DeclareMathAlphabet{\pazocal}{OMS}{zplm}{m}{n}   

\newcommand{\ra}{\rangle}
\newcommand{\la}{\langle}

\newcommand{\one}{\mathbbm 1}
\newcommand*{\at}{@}

\newcommand{\tr}{{\rm Tr}}

\newcommand{\tht}{\tfrac{\theta_1}2}
\newcommand{\pht}{\tfrac{\theta_2}2}

\begin{document}


\title{Neither Weak Nor Strong Entropic Leggett-Garg Inequalities Can Be Violated }\

\author{Christoph Adami}
\email{adami\at msu.edu}
\affiliation{Beyond Center for Fundamental Concepts in Science\\
 Arizona State University, Tempe, AZ 85287}
 \affiliation{Department of Physics and Astronomy, Michigan State University, East Lansing, MI 48824}

\begin{abstract}
The Leggett-Garg inequalities probe the classical-quantum boundary by putting limits on the sum of pairwise correlation functions between classical measurement devices that consecutively measured the same quantum system. The apparent violation of these inequalities by standard quantum measurements has cast doubt on quantum mechanics' ability to consistently describe classical objects. Recent work has concluded that these inequalities cannot be violated by either strong or weak projective measurements~\cite{Adami2019}. Here I consider an entropic version of the Leggett-Garg inequalities that are different from the standard inequalities yet similar in form, and can be defined without reference to any particular observable. I find that the entropic inequalities also  cannot be be violated by strong quantum measurements. The entropic inequalities can be extended to describe weak quantum measurements, and I show that these weak entropic Leggett-Garg inequalities cannot be violated either even though the quantum system remains unprojected, because the inequalities describe the classical measurement devices, not the quantum system. I conclude that quantum mechanics adequately describes classical devices, and that we should be careful not to assume that the classical devices accurately describe the quantum system. 
\end{abstract}

\maketitle

%


\section{Introduction}
The boundary between the classical and quantum realm has never ceased to fascinate. Much focus has been placed on the ability of quantum mechanics to describe both the microscopic and the macroscopic world, since after all we would be surprised to find different theories describing the different realms. Bell's inequalities~\cite{Bell1966} have probed a fundamental aspect of quantum mechanics, namely the non-local nature of the entanglement between spatially separated quantum systems. These inequalities have placed necessary and sufficient conditions for local realism~\cite{Fine1982}, a concept that has now been ruled out by experiment~\cite{FreedmanClauser1972,Aspectetal1982,Weihsetal1998}. The Leggett-Garg inequalities~\cite{LeggettGarg1985}, on the other hand, concern quantum  correlations in time, rather than space. Indeed, the physics of consecutive measurements on the same quantum system promises to hold fascinating insights into the nature of quantum measurement and the concept of physical reality itself. Leggett and Garg formulated their inequalities to test the concept of {\em macrorealism}. Any macrorealistic theory---according to the authors---should abide by these inequalities. According to Leggett and Garg (and many authors subsequently) the inequalities are easily violated by simple quantum measurements, casting doubt on the ability of quantum mechanics to describe both microscopic and macroscopic systems at the same time. 
Numerous experiments appear to have supported this conclusion since then~\cite{Palacios-Laloyetal2010,Gogginetal2011,Xuetal2011,Dresseletal2011,Fedrizzietal2011,Waldherretal2011,Athalyeetal2011,Souzaetal2011,Kneeetal2012,Emaryetal2012,Suzukietal2012,Georgeetal2013,Katiyaretal2013,Asadianetal2014,Robensetal2015,Zhouetal2015,Whiteetal2016,Kneeetal2016} (see also the review~\cite{Emaryetal2014}). 

The conclusions of Leggett and Garg with respect to the apparent violations of their inequalities have not gone uncontested. Ballantine replied with a comment~\cite{Ballantine1987} that stated that a violation of the Leggett-Garg inequalities revealed a contradiction between non-invasive measurability and quantum mechanics, rather than an inability of quantum mechanics to consistently describe macroscopic objects, a view shared among others by Peres~\cite{Peres1989}. Recently I have re-examined this question and showed that Leggett-Garg inequalities cannot be violated in either strong or weak projective measurements~\cite{Adami2019}, and that therefore quantum mechanics consistently describes both microscopic and macroscopic physics. The central observation is similar to Ballantine's, namely that the derivation of the inequalities explicitly assumes that the intermediate of the three measurements is carried out, and that assuming that it is not (as is generally done) violates
the very assumptions behind the derivation. I show this explicitly by considering the  middle measurement to be possibly weak, and showing that even in the limit of a zero-strength (and hence non-existing) measurement, the inequalities cannot be violated. 

Here I consider an alternative form of the Leggett-Garg inequalities, written in terms of quantum entropies~\cite{UshaDevietal2013}. The relation between the entropic Leggett-Garg inequalities and the standard form is precisely the same as the relationship between standard Bell inequalities and their entropic counterpart~\cite{BraunsteinCaves1988} (see also~\cite{CerfAdami1997}). While they appear similar in form, the entropic version of the inequalities actually explores distinct geometric features, and its formulation does not depend on any particular observable. I will first derive entropic Leggett-Garg inequalities using quantum information-theoretic tools I previously employed deriving entropic Bell inequalities~\cite{CerfAdami1997}, and then proceed to show that they cannot be violated by strong measurements. I then formulate extended inequalities that must hold for weak or strong quantum measurements, and show that these cannot be violated either. I then offer some conclusions about the lessons that these inequalities are teaching us about the capacity of quantum mechanics to describe both classical and quantum systems, and the nature of reality.

\section{Entropic Leggett-Garg inequalities for strong measurements}
The standard Leggett-Garg inequalities can be derived simply by insisting that the three measurement devices that consecutively measured an arbitrary quantum state are consistent. For example, for binary devices $A_1$, $A_2$ and $A_3$ that have outcomes $+$ and $-$ and a joint density matrix $\rho_{123}$ that is normalized according to
\be
\sum_{xyz=-}^+\la xyz|\rho_{123}|xyz\ra=1\;,
\ee
the correlation function $K_{12}=\tr( \sigma_z\otimes\sigma_z\rho_{12})$ (for example) between the first two devices is the sum (here, $p(xyz)=\la xyz|\rho_{123}|xyz\ra$)
\be
K_{12}&=&p(---)+p(--+)+p(++-)+p(+++)\nonumber \\
&-&p(+--)-p(+-+)-p(-+-)+p(-++)\;.\ \ \ \ \  \label{sum}
\ee 
Using this expression (and the analogous ones for $K_{23}$ and $K_{12}$) it is easy to show that as long as the $p(xyz)$ are  probabilities, we can immediately derive three inequalities for the correlations
\be
B_1&=&K_{12}+K_{23}-K_{13}\leq1\;, \label{b1}\\
B_2&=&K_{12}+K_{13}-K_{23}\leq1\;,\label{b2}\\
B_3&=&K_{13}+K_{23}-K_{12}\leq1\;, \label{b3}
\ee
which are three of the four standard Leggett-Garg inequalities~\footnote{Another common inequality, written as $K_{12}+K_{13}+K_{23}+1\geq0$ will not be considered here because it has no entropic equivalent.}.

\begin{figure}[htbp] 
   \centering
   \includegraphics[width=3.5in]{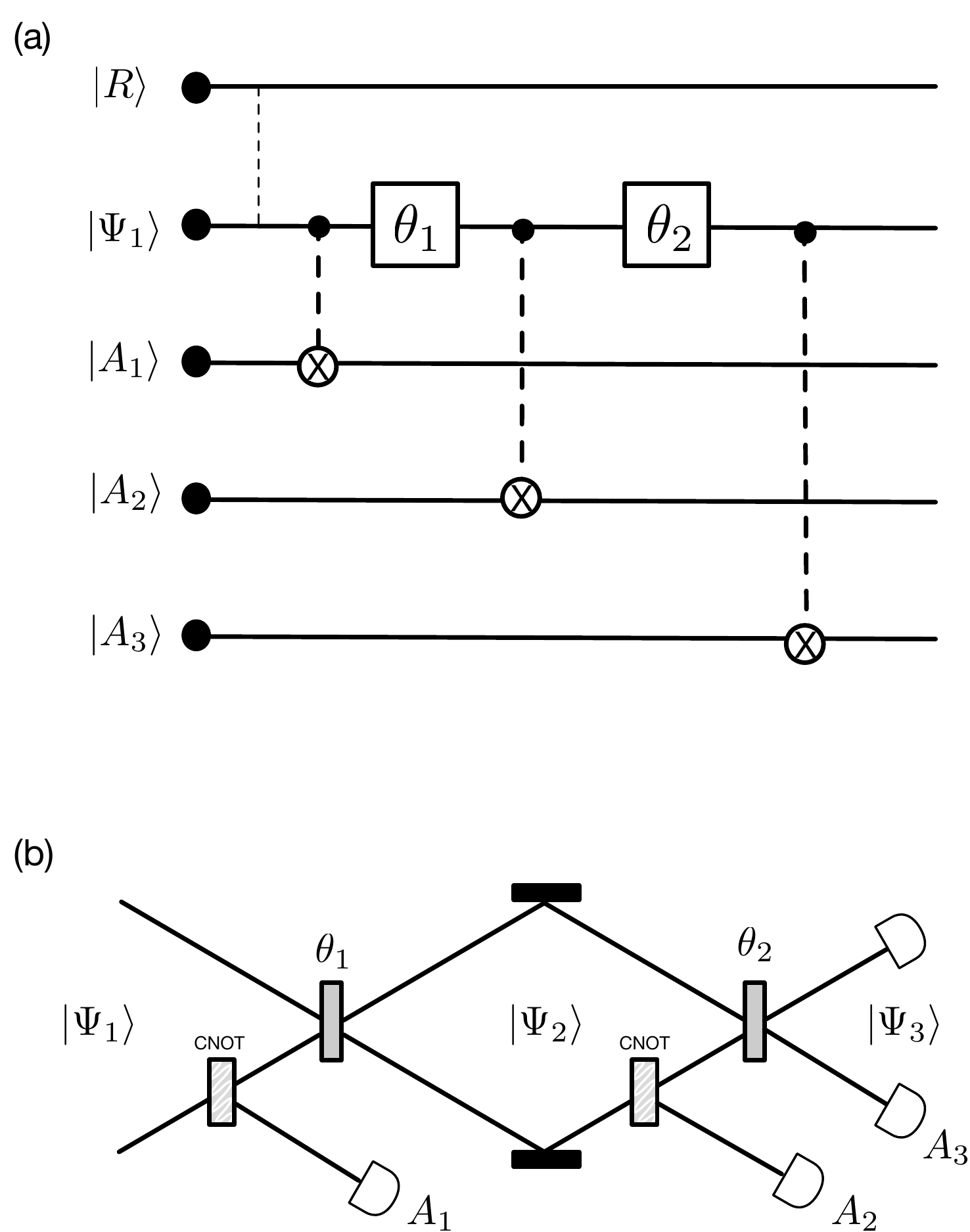} 
   \caption{(a) Three consecutive measurements on an arbitrary quantum state $|\Psi\rangle_1$ (shown entangled with a reference state) $R\ra$ using measurement devices $A_1$, $A_2$, and $A_3$, in terms of a standard circuit diagram. The measurements are performed via controlled-NOT gates. (b) An implementation of three consecutive measurements using a Mach-Zehnder arrangement.}
   \label{fig1}
\end{figure}

To derive entropic inequalities, we first write down an expression for the joint density matrix for the three binary devices $A_1$, $A_2$ and $A_3$ measuring an arbitrary mixed state $\rho$. Fig.~\ref{fig1}a shows the setup in terms of a standard quantum circuit diagram with the input mixed state $\rho$ purified using an arbitrary reference $|R\rangle$. The measurement can also be thought of in terms of a Mach-Zehnder interferometer as in Fig.~\ref{fig1}b, with relative angles $\theta_1$ and $\theta_2$ between the first and second, and the second and third measurement, respectively.  While I will treat the measurement of a random mixed state here, the arguments work just as well in terms of measuring a known prepared state (as in~\cite{Adami2019}) in which case the first measurement can simply be viewed as the state preparation.

\begin{figure}[htbp] 
   \centering
   \includegraphics[width=3in]{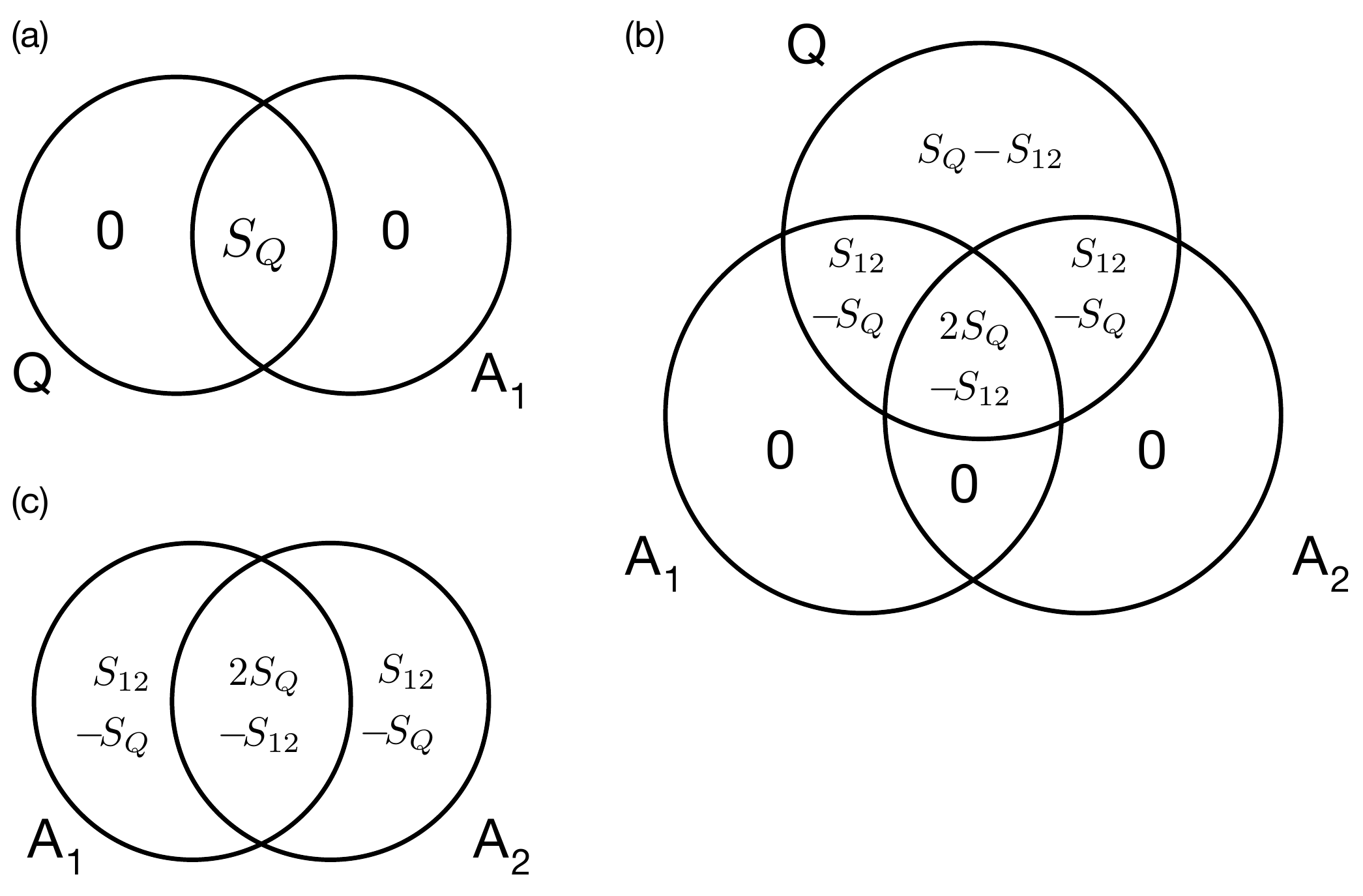} 
   \caption{Entropy Venn diagrams for two ancilla $A_1$ and $A_2$ that consecutively measured the mixed state $\rho_Q$. (a) For strong projective measurements, the entropy of the quantum state is entirely shared with the detector. (b) Joint entropy Venn diagram for the quantum state and two detectors. (c): Venn diagram for the relative state of the two detectors, in terms of the joint entropy $S(A_1A_2)=S_{12}$. }
   \label{fig2}
\end{figure}

\begin{figure}[htbp] 
   \centering
   \includegraphics[width=2in]{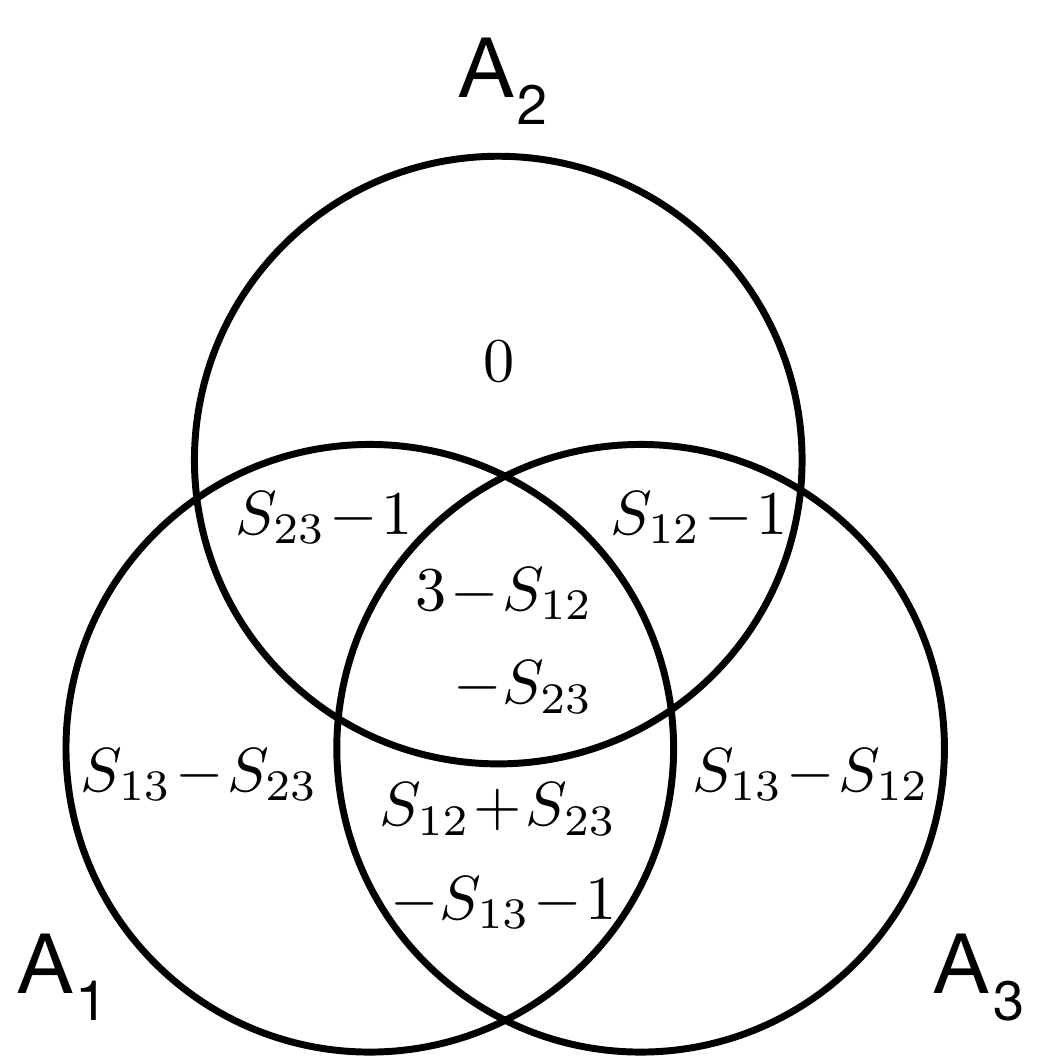} 
   \caption{Entropy Venn diagram for three quantum ancillae that measured a maximally mixed quantum state (from~\cite{GlickAdami2017c}). The entries $S_{12}$, $S_{13}$ and $S_{23}$ are defined in the text.}
   \label{fig3}
\end{figure}

The entropic Leggett-Garg inequalities can be derived from inspecting the entropy Venn diagram for the three measurement devices. Entropy Venn diagrams are  a convenient way to visualize how quantum entropy is distributed among multiple systems. 
Imagine for example a mixed quantum state with density matrix $\rho_Q$ with $S_Q=-\tr\rho_Q\log\rho_Q$ (for qubits, $0\leq S_Q\leq1$).  A strong projective measurement with qubit ancilla $A_1$ will result in the Venn diagram Fig.~\ref{fig2}a, implying that all of the entropy of the quantum state is shared with the detector. If a second detector $A_2$ subsequently measures the same quantum state, its relationship with the quantum state is the same as the relationship depicted in Fig.~\ref{fig2}a. As the entropy of each detector must equal the entropy of the quantum state, the only other variable in the joint Venn diagram Fig.~\ref{fig2}b of the quantum state and two detectors is given by the joint entropy of the two detectors, $S(A_1A_2)\equiv S_{12}$. This value is determined by the relative angle between the two detectors. The two detectors are described by the Venn diagram in Fig.~\ref{fig2}c, which is easily obtained from the one in Fig.~\ref{fig2}b simply by ignoring the quantum state.
If a third detector subsequently measures the same quantum state, the Venn diagram for the relative state of these detectors is depicted in Fig.~\ref{fig3} for the special case where $\rho_Q$ is maximally mixed ($S_Q=1$), as I will assume for simplicity throughout. 
For the fully mixed quantum state $\rho_Q$, the entropy Venn diagram can be written entirely in terms of the pairwise entropies $S_{12}$ as well as $S_{13}$ and $S_{23}$. The peculiar form of this Venn diagram (in which $A_2$ is fully determined given the detectors $A_1$ and $A_3$, that is, the measurements in the past and in the future) is explained in more detail in \cite{GlickAdami2017c}, but in any case I will repeat the calculation below. 

\begin{figure}[htbp] 
   \centering
   \includegraphics[width=2in]{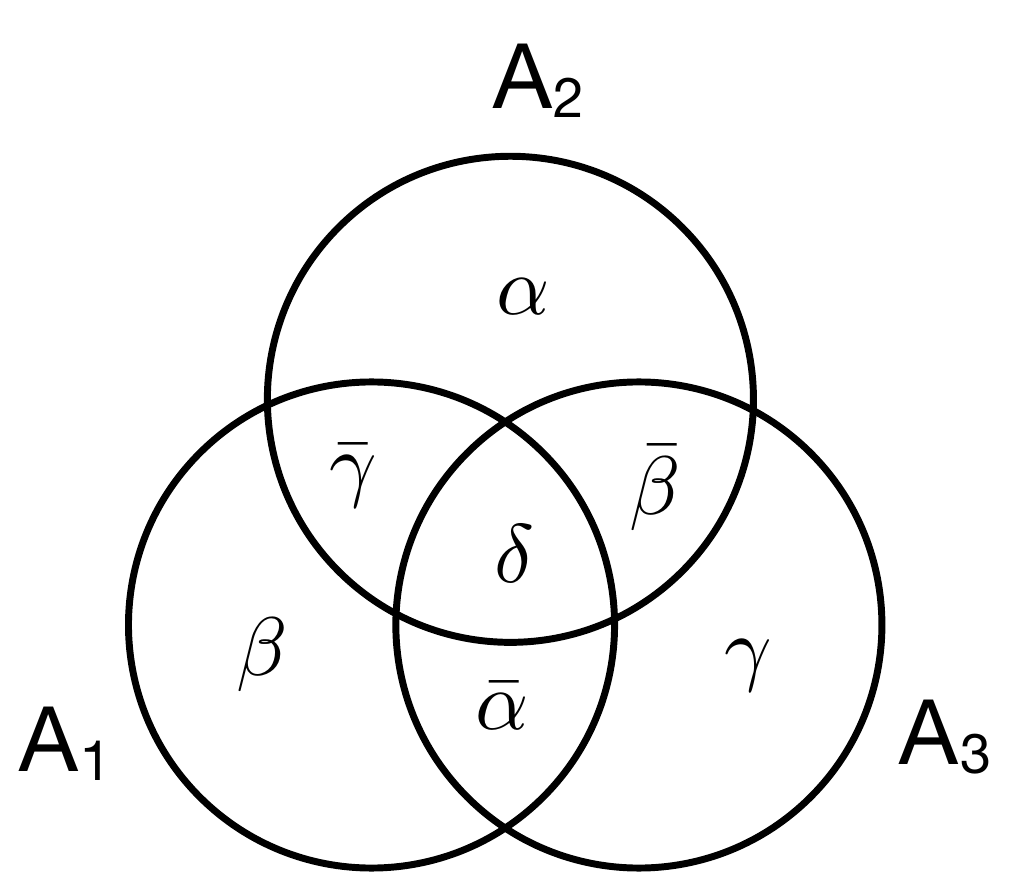} 
   \caption{General tri-partite entropy Venn diagram defining the entries $\alpha$, $\bar\alpha$, $\beta$, $\bar\beta$, $\gamma$, $\bar\gamma$, and $\delta$.}
   \label{fig4}
\end{figure}

It is straightforward to derive entropic inequalities by inspecting the Venn diagram for the three detectors shown in Fig.~\ref{fig3}. Consider first the general tri-partite entropy Venn diagram in Fig.~\ref{fig4}.
In classical (Shannon) theory, all entries except $\delta$ are positive, and in particular the following inequalities that are independent of $\delta$ hold~\cite{CerfAdami1997}
\be
B^\star_1&=&\alpha+\bar\alpha\geq0\;,\\
B^\star_2&=&\beta+\bar\beta\geq0\;,\\
B^\star_3&=&\gamma+\bar\gamma\geq0\;.
\ee
The quantities $\bar\alpha$, $\bar\beta$, and $\bar\gamma$ are positive by themselves due to strong subadditivity. However, in quantum mechanics the conditional entropies $\alpha$, $\beta$, and $\gamma$ can be negative~\cite{CerfAdami1997a}, so the inequalities $B^\star_1$, $B^\star_2$, and $B^\star_3$ could be violated if the system becomes non-classical (as indeed happens in the case of the entropic Bell inequalities), implying that the system is non-separable~\cite{Cerfetal1998a}. By inspection of Fig.~\ref{fig3}, we immediately deduce the entropic Leggett-Garg inequalities
\be
B^\star_1&=&S_{12}+S_{23}-S_{13}\geq1\;, \label{bstar1}\\
B^\star_2&=&S_{12}+S_{13}-S_{23}\geq1\;,\\
B^\star_3&=&S_{13}+S_{23}-S_{12}\geq1\;,
\ee
in remarkable analogy to Eqs.~(\ref{b1}-\ref{b3}). Note that $B_1^\star$ is identical with the entropic Leggett-Garg inequality proposed in Ref.~\cite{UshaDevietal2013}, which they derived using the chain rule and imposing subadditivity. We will see below that indeed only $B_1^\star$ imposes strong limits, while $B_2^\star$ and  $B_3^\star$ are trivially obeyed. 

I will now show explicitly that $B_1^\star$ cannot be violated with strong measurements, contrary to the claim in~\cite{UshaDevietal2013} and 
\cite{Katiyaretal2013}. In a sense, this is immediately obvious using the Venn diagram formulation because $B_1^\star$ is guaranteed to be positive owing to strong subadditivity (since $\alpha=0$). However, doing the explicit calculation will reveal the discrepancy more clearly. 

The calculation I develop here follows the formalism developed in Ref.~\cite{GlickAdami2017c}. Using qubits as detectors is convenient, but we could have just as easily used any multiple-state detector with eigenstates $|x\ra$, in which case the Hadamard operator of the CNOT gate ($\sigma_x$, see below) would be replaced with translation operators $T$ so that $T(x)|0\ra=|x\ra$ (see, for example,~\cite{Curicetal2018}).
We can assume that an (unknown) qubit can be written in terms of the orthogonal basis states $|Q\rangle$ and $|\bar Q\rangle $ as well as reference states $|R\rangle$ and $|\bar R\rangle$ as 
\be
|\Psi\ra=\frac1{\sqrt{2}}(|QR\rangle +|\bar Q\bar R\rangle)\;
\ee 
so that $\rho_Q=\tr_R |\Psi\ra\la\Psi|$ is a maximally mixed state with unit entropy. 
The reference states $R$ can be thought of as representing all the states that conceivably had interacted with the quantum system in the past and are not, in general, experimentally accessible.

I will write the orthogonal (first) qubit ancilla states as $|0\rangle_1$ and $|1\rangle_1$ and assume (without loss of generality) that they refer to horizontal and vertical polarizations, respectively (this is at the experimenter's discretion). The first measurement is implemented using the unitary operator that measures the quantum state in the ($H,V$) basis (again, without loss of generality)
\be
U_1=|H\rangle\langle H|\otimes \one +|V\ra\la V|\otimes \sigma_x\;,
\ee 
where $\sigma_x$ (the first Pauli matrix) flips the polarization of the qubit ancilla and we can assume that $|H\ra\la H|Q\ra=|H\ra$ etc. because the arbitrary state $|Q\ra$ must be arbitrary in any basis. After the first measurement, we are left with the wave function
\be
|\Psi_1\ra=\frac1{\sqrt2}(|HR\ra|0\ra_1+|V\bar R\ra|1\ra_1\;,
\ee
The first detector then has the density matrix $\rho_1=\frac12(|0\ra_1\la0|+|1\ra_1\la1|)$ and entropy $S(A_1)=-\tr(  \rho_1\log_2\rho_1) \equiv S_1=1$.

As advertised, the second measurement will be performed at an angle $\theta_1$ with respect to the first, using the basis states $|\theta\ra_1$ and $|\bar \theta\ra_1$.
To measure in this basis, rewrite the quantum system's basis states in the $\theta_1$-basis:
\be
|H\ra&=&\cos(\tht)|\theta_1\ra-\sin(\tht)|\bar\theta_1\ra\\
|V\ra&=&\sin(\tht)|\theta_1\ra+\cos(\tht)|\bar\theta_1\ra\;.
\ee
which defines $|\theta_1\ra$ and $|\bar\theta_1\ra$ implicitly. 
The (strong) measurement operator can then be written as
\be
U_2=|\theta\ra\la\theta|\otimes\one +  |\bar\theta\ra\la\bar\theta|\otimes \sigma_x
\ee
where $\sigma_x$ acts on the second qubit ancilla. This measurement is invasive (as must be all quantum measurements, see for example Peres' discussion of this issue in~\cite{Peres1989}), but we will consider less invasive measurements using the weak measurement paradigm in the following section. The wave function after the second measurement becomes 
\begin{widetext}
\be
|\Psi_2\ra=U_2|\Psi_1\ra|0\ra_2=\frac1{\sqrt2}\biggl(|\theta_1\ra\Bigl[\cos(\tht)|R\ra|0\ra_1-\sin(\tht)|\bar R\ra|1\ra_1\Bigr]|0\ra_2+|\bar\theta_1\ra\Bigl[\cos(\tht)|\bar R\ra|1\ra_1-\sin(\tht)|R\ra|0\ra_1\Bigr]|1\ra_2\biggr)\; \; \; \;.
\ee
\end{widetext}
The joint density matrix for the two devices is diagonal in the pointer basis (or, as we say, ``classical")
\begin{widetext}
\be
\rho_{12} &=& \tr_{QR}(|\Psi\ra_2\la\Psi|)=\frac12\biggl(|0\ra_1\la0|\Bigl[\cos^2(\tht)|0\ra_2\la0|+\sin^2(\tht)|1\ra_2\la1|\Bigr]+|1\ra_1\la1|\Bigl[\sin^2(\tht)|0\ra_2\la0|+\cos^2(\tht)|1\ra_2\la1|\Bigr]\nonumber\\
&=&  \frac12  \begin{pmatrix}
      \cos^2(\tht) & 0 & 0 & 0 \\
      0 & \sin^2(\tht) & 0 & 0\\
      0& 0 & \sin^2(\tht) & 0 \\
      0 & 0 & 0 &  \cos^2(\tht)
   \end{pmatrix}\;.
   \ee
\end{widetext}
\begin{figure}[htbp] 
   \centering
   \includegraphics[width=2.5in]{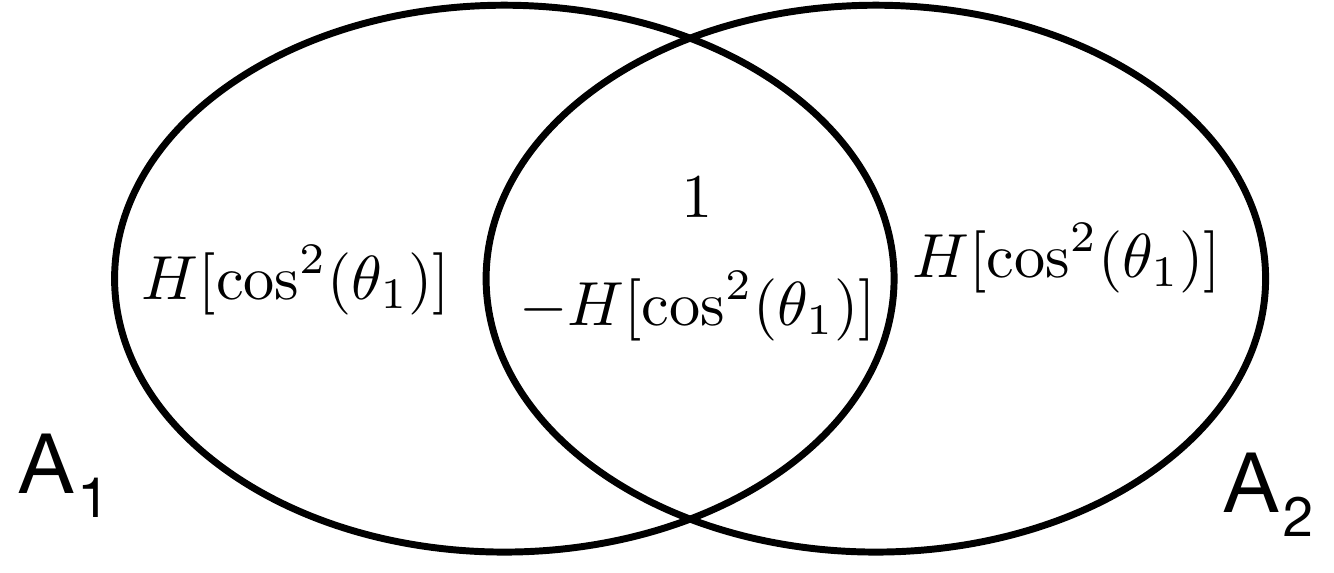} 
   \caption{Bipartite Venn diagram for the first two detectors, where $S_{12}=1+H[\cos^2(\theta_1)]$}
   \label{fig5}
\end{figure}
Using $\rho_{12}$ we can calculate the joint entropy between detectors $A_1$ and $A_2$ by evaluating $S_{12}=-\tr\rho_{12}\log\rho_{12}$
\be
S_{12}=1+H[\cos^2(\theta_1)]\;. \label{entzero}
\ee
where $H[x]=-x\log_2 x-(1-x)\log_2(1-x)$ is the binary entropy function. Equation (\ref{entzero}) is the standard result, and corresponds to the correlation coefficient between the first and second detectors $K_{12}=\cos(\theta_1)$. We can also use this result to fill in the Venn diagram in Fig.~\ref{fig2}c, which I show in Fig.~\ref{fig5}.

We are now ready to perform the third (strong) measurement, with the third device set an angle $\theta_2$ with respect to the second one. To do this, rewrite the quantum system's basis states, currently written in the $\theta_1$-basis, in the $\theta_2$-basis instead:
\be
|\theta_1\ra&=&\cos(\pht)|\theta_2\ra-\sin(\pht)|\bar\theta_2\ra\;,\\
|\bar\theta_1\ra&=&\sin(\pht)|\theta_2\ra+\cos(\pht)|\bar\theta_2\ra\;.
\ee

With a third qubit ancilla's basis states $|0\ra_3$ and $|1\ra_3$, the wave function after the third measurement is obtained using the unitary operator $U_3=|\theta_2\ra\la\theta_2|\otimes\one+|\bar\theta_2\ra\la\bar\theta_2|\otimes \sigma_x$, with $\sigma_x$ acting on the third qubit's Hilbert space, reads
\begin{widetext}
\be
|\Psi_3\ra=\frac1{\sqrt2}\Biggl\{|\theta_2\ra&\biggl(&\cos(\pht)\Bigl[\cos(\tht)|R\ra|0\ra_1+\sin(\tht)|\bar R\ra|1\ra_1\Bigr]|0\ra_2+\sin(\pht)\Bigl[\cos(\tht)|\bar R\ra|1\ra_1-\sin(\tht)|R\ra|0\ra_1\Bigr]|1\ra_2\biggr)|0\ra_3 \nonumber\\
+|\bar\theta_2\ra&\biggl(&\cos(\pht)\Bigl[\cos(\tht)|\bar R\ra|1\ra_1-\sin(\tht)|R\ra|0\ra_1\Bigr]|1\ra_2-\sin(\pht)\Bigl[\cos(\tht)|R\ra|0\ra_1+\sin(\tht)|\bar R\ra|1\ra_1\Bigr]|0\ra_2\biggr)|1\ra_3\biggr\}\;.\;\;\;\; \; \; \;  \label{psi3}
\ee
\end{widetext}
The joint density matrix of three devices $\rho_{123}=\tr_{QR}(|\Psi_3\ra\la\Psi_3|)$ then becomes (here, I abbreviate $c_1\equiv \cos(\tht)$, $c_2\equiv \cos(\pht)$, $s_1\equiv \sin(\tht)$, $s_2\equiv \sin(\pht)$)
\begin{widetext}
\be
\rho_{123}=\frac12
\begin{pmatrix}
      c_1^2 c_2^2 & 0 & -s_1c_1s_2c_2& 0 &0 &0& 0 &0 \\
      0 & c_1^2 s_2^2 & 0 &  s_1c_1s_2c_2 & 0 & 0 & 0 & 0\\
        -s_1c_1s_2c_2& 0 &s_1^2 s_2^2& 0 & 0 & 0 & 0 & 0  \\
      0 & s_1c_1s_2c_2 & 0 &s_1^2c_2^2 & 0 & 0 & 0 & 0 \\
      0 & 0 & 0 & 0 & s_1^2c_2^2 & 0 & s_1c_1s_2c_2 & 0 \\
      0 & 0 & 0 & 0 & 0 & s_1^2 s_2^2 & 0 & -s_1c_1s_2c_2 \\
      0 & 0 & 0 & 0 & s_1c_1s_2c_2 & 0 & c_1^2 s_2^2 & 0 \\
      0 & 0 & 0 & 0 & 0 & -s_1c_1s_2c_2 & 0 & c_1^2 c_2^2
   \end{pmatrix}\; .\; \; \; \; \; \;  \label{123}
   \ee
   \end{widetext}
The pairwise matrix $\rho_{23}$ is obtained by tracing out the first device, with entropy
\be 
S_{23}=-\tr \rho_{23}\log \rho_{23}= 1+H[\cos^2(\theta_2)]\;,
\ee
as expected. Tracing over the middle detector will reveal the difference to the literature (where the second device is assmed {\em not} to be measured), giving instead
\be
S_{13}=1+H\Bigl[\cos^2(\tht)\cos^2(\pht)+\sin^2(\tht)\sin^2(\pht)\Bigr]\;. \label{middle}
\ee
Assuming that the middle measurement does not take place would have instead yielded $S_{13}=1+H[\cos^2(\theta_1+\theta_2)]$ (I will show  in more detail below what happens when we perform the middle measurement weakly). However, that result is completely incompatible with the density matrix (\ref{123}), which after all is the main ingredient in the derivation of standard or entropic Leggett-Garg inequalities. 
This observation parallels what I found for standard Leggett-Garg inequalities, where the correct correlation function between the first and third device is $K_{13}=\cos(\theta_1)\cos(\theta_2)$~\cite{Adami2019}, a result that incidentally can also be read off straight from Eq.~(\ref{123}) and the expression for $K_{13}$ corresponding to (\ref{sum}). 

We can now check inequality $B_1^\star$. Assuming for simplicity $\theta_1=\theta_2\equiv\theta$ and defining $H_{13}\equiv H\Bigl[\cos^4(\tfrac{\theta}2)+\sin^4(\tfrac{\theta}2)\Bigr]$ the entropic Leggett-Garg inequalities become
\be
B^\star_1&=&2H[\cos^2(\theta)]-H_{13}\geq0\;, \label{b1h}\\
B^\star_2&=&B^\star_3=H_{13}\geq0\;.
\ee
It is immediately clear (from using Jensen's inequality) that the entropic Leggett-Garg inequality $B^\star_1$ cannot be violated for strong measurements, just as its non-entropic counterpart~\cite{Adami2019}.  Of course, $B^\star_2$ and $B^\star_3$ also cannot be violated by either weak or strong measurements, as the Shannon entropy $H_{13}$ is positive definite.

\section{Leggett-Garg inequalities for weak measurements}

To implement a weak as opposed to strong measurement in the ``middle" position (the measurement performed by $A_2$), the second ancilla needs to be moved by a measurement from its preparation $|0\ra_2$ not into the orthogonal state $|1\ra_2$ via flipping, but instead the weak measurement should move it only by a small angle to
\be
|\epsilon\ra_2=\sqrt{1-\epsilon^2}|0\ra_2+\epsilon|1\ra_2\;,
\ee
where $0\leq\epsilon\leq1$ parameterizes the strength of the measurement.
This can be implemented simply by using a unitary von Neumann measurement operator $U_2=e^{-iH_\theta}$ with the interaction Hamiltonian $H_\theta=gP_\theta\otimes \sigma_y$ with $P_\theta=|\theta\ra\la\theta|$ and where $\cos(g)=\sqrt{1-\epsilon^2}$~\cite{Aharonovetal1988,Lundeenetal2011,LundeenBamber2012,Dresseletal2014,Curic2018}. Clearly, the strong measurement returns in the limit $g\to\pi/2$ $(\epsilon\to1)$. In the limit $\epsilon\to0$, no measurement takes place.

\begin{figure}[htbp] 
   \centering
   \includegraphics[width=3.5in]{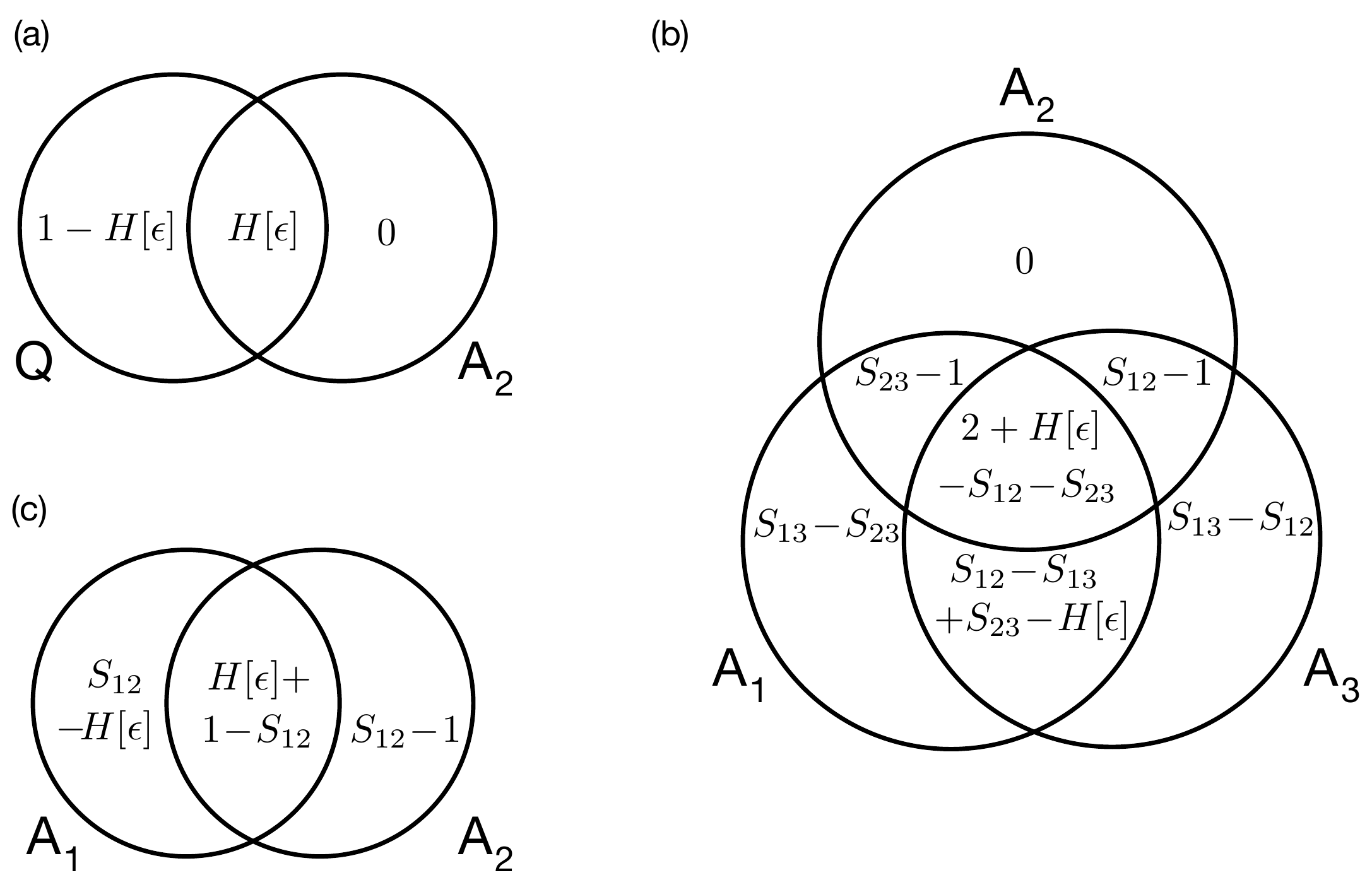} 
   \caption{(a) A weak projective measurement by $A_2$ does not fully resolve the maximally mixed quantum state, leaving $1-H[\epsilon]$ of its entropy unshared. (b) The tri-partite quantum Venn diagram of the three detectors is modified when the middle measurement is weak. Fig.~\ref{fig3} returns in the limit of $\epsilon\to1$ as in that limit $H[\epsilon]\to1$. (c) The relative state of the first two detectors when the second one is weak. In the limit $\epsilon\to0$, $A_1$ and $A_2$ are uncorrelated. }
   \label{fig6}
\end{figure}

Implementing weak measurements results in a wave function $|\Psi\ra_3$ just like (\ref{psi3}) but with $|1\ra_2$ replaced with $|\epsilon\ra_2$ wherever it occurs. The resulting pairwise entropies are the following
\be
S_{12}&=&1+H\Bigl[\frac12+\frac12\sqrt{1-4\epsilon^2\sin^2(\tht)\cos^2(\tht)}\Bigr]\;,\\
S_{23}&=&1+H\Bigl[\frac12+\frac12\sqrt{1-4\epsilon^2\sin^2(\pht)\cos^2(\pht)}\Bigr]\;
\ee
and
\begin{widetext}
\be
S_{13}=1+H\Bigl[\cos^2(\tht)\cos^2(\pht)+\sin^2(\tht)\sin^2(\pht)-2\sqrt{1-\epsilon^2}\sin(\tht)\cos(\tht)\sin(\pht)\cos(\pht)\Bigr]\;. \label{middle}
\ee
\end{widetext}
Using these entropies in (\ref{bstar1}) would suggest that weak measurements can certainly violate the entropic Leggett-Garg inequality $B_1^\star$. Indeed, using again $\theta_1=\theta_2=\theta$, the equation analogous to (\ref{b1h}) can be violated when using $H_{13}=H[\cos^4(\theta/2)+\sin^4(\theta/2) -2\sqrt{1-\epsilon^2}\sin^2(\theta/2)\cos^2(\theta/2)]$. However, this apparent violation is due to the fact that in weak measurements, the entropic Leggett-Garg inequality must be modified. When measurements are weak ($\epsilon<1$), the entropy of the weak device is not fully correlated with the quantum state, and is therefore less than one. This is verified by calculating the entropy directly, giving
$S(A_2)=H[\frac12(1+\sqrt{1-\epsilon^2})]$, using again the binary entropy function defined above. Fig.~\ref{fig6}a shows that under a weak measurement, only part of the entropy of the quantum state is shared with the device, and in the limit $\epsilon\to0$, none of it. 
The tri-partite Venn diagram thus must be modified accordingly, and is shown in Fig.~\ref{fig6}b. Finally, Fig.~\ref{fig6}c shows how a weak measurement using $A_2$ (while $A_1$ and $A_3$ are measuring strongly) changes the relative state between $A_1$ and $A_2$. In the limit of a zero-strength weak measurement (that is, not performing the $A_2$ measurement) $A_2$ is completely decoupled from the quantum system and all other detectors as $S_{12}\to1$. 
The entropic Leggett-Garg inequality for weak intermediate measurement can be read off the Venn diagram in Fig.~\ref{fig6}b as
\be
B_1'(\epsilon)=S_{12}-S_{13}+S_{23} - H[\epsilon]\geq0\;,
\ee
and naturally it cannot be violated because $B_1'(\epsilon)$ is guaranteed to be positive owing to strong subadditivity of quantum entropies. 

Finally, let us take a look at the relative state of $A_1$ and $A_3$ when the $A_2$ measurement is not performed. Recall that the literature has generally assumed that a ``non-invasive" measurement  (such as for example via the ``ideal negative result" technique,~\cite{Kneeetal2012}) is tantamount to not performing the measurement. Indeed, in the limit $\epsilon\to0$ we  find for the entropy $S_{13}$ [confer Eq.~(\ref{middle})]
\be
S_{13}\stackrel{\epsilon\to0}{\longrightarrow}1+H[\cos^2(\theta_1+\theta_2)]\;.
\ee
But in this limit, the Leggett-Garg inequalities become trivial, as $S_{12}$ and $S_{23}$ both tend toward one. We thus encounter precisely the same situation that we witnessed when analyzing the standard Leggett-Garg inequalities~\cite{Adami2019}: assuming that the middle measurement is not performed when calculating $K_{13}$ or $S_{13}$ enforces vanishing $K_{12}$ ($S_{12}=1$). It is nor permissible to use the expressions from a separate experiment where the $A_2$ measurement was actually performed, as is done in the majority of experiments listed in~\cite{Palacios-Laloyetal2010,Gogginetal2011,Xuetal2011,Dresseletal2011,Fedrizzietal2011,Waldherretal2011,Athalyeetal2011,Souzaetal2011,Kneeetal2012,Emaryetal2012,Suzukietal2012,Georgeetal2013,Katiyaretal2013,Asadianetal2014,Robensetal2015,Zhouetal2015,Whiteetal2016,Kneeetal2016}. Furthermore, an ``ideal negative result" experiment is not a weak measurement at all, and will disturb the quantum wavefunction just as much as a standard projective measurement, as pointed out long ago by Dicke~\cite{Dicke1981}. For those experiments where the ``ideal negative result" technique was used by discarding the half of experiments where the detector clicked, this can be tested directly by using instead the discarded half to calculate the Leggett-Garg inequalities. The analysis will reveal that nothing changed: it does not matter whether the detector clicked or not, as the detector's state is not diagnostic of the quantum state.

Some might argue that writing down ``three-point" Leggett-Garg inequalities (that is, where a measurement is performed at all three points) is futile because once the second measurement is performed the quantum state is ``destroyed" and cannot be measured again. However, this is not true, and careful experimentation can measure the quantum state repeatedly. For example, Katiyar et al.~\cite{Katiyaretal2013} actually measured the three-point function [the diagonal of Eq.~(\ref{123})] using nuclear magnetic resonance techniques, only to find that they were unable to reproduce (via marginalization) the two-point function obtained when not performing the intermediate measurement. They remarkably concluded that ``the grand probability is not legitimate in this case"~\cite{Katiyaretal2013}, when in fact it is the two-point functions that are not legitimate.  Had they used the full $P(x,y,z)$ for the calculation of their entropic Leggett-Garg inequality, they would have found it inviolate. Other approaches using quantum optics also make it possible to perform multiple measurement of the same quantum state in a row, as long as the experimenter does not expect that there is no backreaction on the quantum system~\cite{Curicetal2018}. These backreactions are the essence of quantum mechanics, and are fully accounted for in the Leggett-Garg inequalities.

\section{Discussion}

One of the more remarkable aspects of the entropic version of the Leggett-Garg inequalities is that there is no reference to macroscopic realism or non-invasiveness of measurements in its derivation, two criteria that are usually held as the foundations of the Leggett-Garg inequalities. However, previous work has implied that non-invasiveness is not a necessary criterion for writing down such inequalities~\cite{Waldherretal2011} (see also the discussion in~\cite{Emaryetal2014}). After all, the inequalities follow simply from consistency of the joint density matrix of the measurement devices, and indeed I have argued before~\cite{Adami2019} that non-invasive quantum measurements can only exist if a quantum state is measured in the basis in which it was prepared, which is really a classical measurement. 
I suspect that using non-projective quantum measurements (for example, POVMs) does not alter any of the conclusions offered here. For example, one way to implement weak measurements is to perform projective measurements in an extended Hilbert space. As weak measurements cannot violate the Leggett-Garg inequalities, I suspect that POVMs cannot either, but that proof is still outstanding. 

The widespread and ubiquitous apparent violation of Leggett-Garg inequalities has caused considerable headscratching among researchers, because ostensibly the violation implies that either quantum mechanics does not correctly describe macroscopic objects (if macrorealism is true), or else we should experience violation of macrorealism in everyday life (if quantum mechanics is correct instead)~\cite{Leggett2002}. Today there is plenty of evidence that the superposition of macroscopic objects is not only real, but can be experimentally verified (see, for example,~\cite{Kovachyetal2015}). Why then do we do not experience this in everyday life? Kofler and Brukner \cite{KoflerBrukner2008} offer two possible escapes from this conundrum: either microscopically distinct states (what we call ``orthogonal states") do not actually exist in nature (meaning, all measurements are necessarily a little bit weak) or else decoherence destroys the violations before they could ever be experienced. Here I suggest that instead there is no paradox because the Leggett-Garg inequalities (entropic or not) are never violated. While quantum superpositions of macroscopic objects are real, we cannot experience them because measurement devices cannot reflect such a state of affairs. The central misunderstanding then is not that quantum mechanics does not adequately describe macroscopic objects. The central misunderstanding is the belief that classical measurement devices describe the quantum objects that they are designed to reflect. It is natural, in classical physics, to assume that a measurement device allows you to infer the state of the measured system, as this is precisely the role of the measurement device. But in quantum mechanics, this is provably no longer the case (in fact, this is the essence of the quantum no-cloning theorem). Measurement devices allow us to infer the state of other measurement devices, but (in the worst case), are completely unreliable in inferring the state of the quantum system. The ``ideal negative result" measurement (also called ``interaction-free" measurement) setups are a case in point. The absence of a click lulls some of us into believing that the quantum system ``is" in the state that the absent click would indicate. But the device has lied: the silence means nothing. 

It is of course anti-climactic to re-analyze experiments to show that Leggett-Garg inequalities cannot be violated. However, I believe it is worth while carrying those out, because even though those inequalities do not test macrorealism, they do throw light on the physics of consecutive quantum measurements and the relation between the classical and quantum description of the world~\cite{GlickAdami2017}. The entropic version of the inequalities are particularly interesting because they are able to probe the non-diagonal entries in the full density matrix (\ref{123}). While those entries play no role in the pair-wise marginal density matrices when measurements are strong, sophisticated tomography techniques can reveal them in the full density matrix~\cite{Curicetal2019}. In particular, there may exist extended entropic inequalities involving double-conditional entropies [such as $S(A_3|A_2A_1)$] that can be violated, to show experimentally that quantum measurements are non-Markovian, as theory suggests~\cite{GlickAdami2017c}. 

\begin{acknowledgments}
I would like to acknowledge the hospitality of Paul Davies and the Beyond Center for Fundamental Concepts in Science at Arizona State University, where this research was carried out. I also would like to thank Paul Davies and G. Andrew Briggs for discussions.  This work was supported in part by a grant from the John Templeton Foundation.
\end{acknowledgments}

\bibliography{quant-PRA}

\end{document}